\documentclass[12pt]{article}

\usepackage{amsmath,amssymb}
\usepackage{graphicx}
\makeatletter

\@addtoreset{equation}{section}
\makeatother

\def\beq#1\eeq{\begin{align}#1\end{align}}

\newcommand{\scZ}{\ensuremath{\mathcal{Z}}}

\begin{document}

\baselineskip=18pt  

\begin{titlepage}

\setcounter{page}{0}

\renewcommand{\thefootnote}{\fnsymbol{footnote}}

\begin{flushright}
CALT-68-2763\\
IPMU09-0142\\
UT-09-26
\end{flushright}
\vskip 2.5cm

\begin{center}
{\LARGE \bf
Open BPS Wall Crossing
and
M-theory
}

\vskip 1.2cm

{\large
Mina Aganagic$^{1,2}$ and
Masahito Yamazaki$^{2,3,4}$
}

\vskip 0.8cm

{
\it
$^1$Center for Theoretical Physics, University of California, Berkeley, CA 94720, USA\\

$^2$IPMU,
University of Tokyo, 
Chiba 277-8586,
Japan\\

$^3$Department of Physics, University of Tokyo,
Tokyo 113-0033,
Japan\\

$^4$California Institute of Technology, 
CA 91125, USA\\
}

\end{center}

\vspace{1.2cm}

\centerline{{\bf Abstract}}
\medskip
\noindent

Consider the degeneracies of BPS bound states of one D6 brane wrapping Calabi-Yau $X$ with D0 branes and D2 branes. When we include D4-branes wrapping Lagrangian cycle $L$ in addition, D2-branes can end on them. These give rise to new bound states in the $d=2$, ${\cal N}=(2,2)$ theory of the D4 branes. We call these ``open" BPS states, in contrast to closed BPS states that arise from D-branes without boundaries. Lifting this to M-theory, we show that the generating function is captured by free Fock space spanned by M2-brane particles ending on M5 branes wrapping $L$.
This implies that the open BPS bound states are counted by the square of the open topological string partition function on $X$, reduced to the corresponding chamber.
Our results give new predictions for open BPS invariants and their wall crossing phenomena when we change the open and closed string moduli.
We relate our results to the work of Cecotti and Vafa on wall crossing in the two dimensional ${\cal N}=(2,2)$ theories. The findings from the crystal melting model for the open BPS invariants proposed recently fit well with the M-theory predictions.

\end{titlepage}
\setcounter{page}{1} 


\section{Introduction}

Topological string theory has proven to be a powerful tool for counting of BPS states in various contexts. Closed topological string theory on $X$ is counting BPS states in the 5d theory resulting from compactifying M-theory on $X$. These are spinning M2 branes wrapping curves on $X$, where one considers a gas of such particles. It is also computing the degeneracies of BPS bound states in the 4d ${\cal N}=2$ theory from compactifying IIA on $X$. These are bound states of D0 and D2 branes with a single D6 brane wrapping $X$.
The relation of these two counting problems was derived in \cite{DVVafa}, via lifting IIA to M-theory. There is a subtlety in the problem of counting BPS bound states due to wall crossing phenomena --- the BPS degeneracies depend on the moduli at infinity.
In \cite{AOVY} we showed that, at least when the Calabi-Yau contains no 4-cycles, the topological string is computing also the wall crossing of the D6-D2-D0 degeneracies.
We showed that there is a simple formula for the BPS generating function
\beq
\scZ^c_{\rm BPS}(q,Q)=\scZ^c_{\rm top}(q,Q) \scZ^c_{\rm top}(q,Q^{-1}) \Big|_{\rm chamber},
\label{eq1}
\eeq
which states that BPS partition function is given as a reduction of the square of the closed topological string partition function.

In addition to a D6 brane on $X$, we can also include D4-branes wrapping special Lagrangian submanifolds of $X$. The BPS states in the resulting ${\cal N}=(2,2)$ theory in two dimensions are ``open" BPS states. These are D2 branes wrapping holomorphic disks and ending on the D4 branes. In this paper, we generalize the argument of \cite{DVVafa} to the case of the open BPS bound states. We show that they are naturally counted by the three-dimensional ${\cal N}=2$ theory corresponding to M-theory on $X$ with M5 branes wrapping Lagrangian submanifold $L$ of $X$. The BPS states are a gas of spinning M2 branes, ending on the M5 branes, in three dimensions. The latter counting problem is in turn solved by the open
topological string. The topological string theory also captures the
chamber dependence\footnote{Chamber dependence of open BPS states, in the context of surface operators of ${\cal N}=2$, $d=4$ gauge theories, was discussed recently in \cite{GaiottoSurface}.}, as we change the open and closed string moduli.

In fact, from the Calabi-Yau perspective, it is natural to consider all BPS particles together, open and closed. Let us denote by  $\scZ_{\rm BPS}$ the corresponding partition function. Then M-theory lift predict that the chamber dependence
of this is captured by the open+closed  partition function as
\beq\label{wc}
\scZ^{o+c}_{\rm BPS}(q,Q,v)=\scZ^{o+c}_{\rm top}(q,Q,v) \scZ^{o+c.*}_{\rm top}(q,Q^{-1},v^{-1}) \Big|_{\rm chamber},
\eeq
where $\scZ^{o+c.*}_{\rm top}(q,Q^{-1},v^{-1})$, whose closed string part coincides with $\scZ^c_{\rm top}(q,Q^{-1})$ in \eqref{eq1}, will be defined in the main text. 

The BPS spectra of two dimensional ${\cal N}=(2,2)$ theory were studied previously by Cecotti and Vafa \cite{CecottiVold}. In fact, \cite{CecottiVold} were first to discover the discontinuities in the BPS spectra as the moduli of the theory are varied. We show that M-theory results for the jumps in the BPS spectrum of D6-D2-D0 branes agree, up to some subtleties, with the general predictions of \cite{CecottiVold}.

The organization of the present paper is as follows. In section \ref{sec.closedwc}, we review the basic idea of \cite{AOVY}. In section \ref{sec.opentop} we review the open topological string partition function, focusing on the integrality  \cite{OVknot,LMV} and the connection with counting M2-branes ending on M5-branes. We also generalize the work of \cite{DVVafa} to open topological string and open BPS bound states.
Based on these results we derive in section \ref{sec.openwc} the wall crossing of the open BPS partition function. In section 4 we discuss the relation to the work of Cecotti and Vafa.
\\

\noindent{\bf Note added}

During the preparation of this paper we learned that related results on open BPS invariants have been obtained independently in \cite{DSV}.
We thank R.~ Dijkgraaf, P.~ Sulkowski and C.~ Vafa for sharing this information with us.


\section{Closed BPS Wall Crossing and M-theory}\label{sec.closedwc}

The basic idea behind \cite{AOVY} is simple.
Consider type IIA string theory on a non-compact Calabi-Yau manifold $X$, with a D6 brane wrapping $X$.
This can form BPS bound states with D0 and D2 branes wrapping holomorphic curves.
The IIA on $X$ in the presence of the D6 brane is dual to M-theory on $X \times TN \times R$, where $TN$ is the
single-centered Taub-NUT space. Taub-NUT space is a $S^1$ fibration over $R^3$, and the $S^1$ has constant radius at infinity. Let us denote the radius by $R$. The D0-branes are mapped to KK momenta on the TN circle, and D2-branes to M2-branes wrapping curves in $X$.

The counting problem of M-theory on $X\times TN\times R$ is closely related to the corresponding counting problem of M-theory on $X\times R^{4,1}$.  Each single particle state in M-theory on $X\times R^{4,1}$ gives a single particle state in M-theory on $X\times TN\times R$. However, we also need to consider multi-particle states on $TN$, corresponding to multiple D-branes bound to the D6 branes.
Under two assumptions, this is captured by a gas of free particles on $X\times R^{4,1}$. First, assume that there $X$ has no compact 4-cycles. This means that in IIA there is no D4-branes wrapping 4-cycles, or in M-theory there is no M5-brane wrapping $S^1$ of the Taub-NUT, so that on $R^{4,1}$ we need to consider only particles, and not strings.
The second assumption is that we can choose the value of the moduli such that all the M2-brane particles preserve the same supersymmetry and there is no forces acting between them.
When the first assumption is satisfied, the second one is true as well. We can start wit a large Fock space, containing all possible M2 brane and anti-M2 brane states of the theory on $X\times R^{4,1}$. To get the BPS states in a particular chamber,
we restrict to the Fock space of 1-particle M2 brane states whose central charges are aligned, and which preserve the same supersymmetry. The partition function of BPS bound states of the D6 brane with these branes is simply the trace in the restricted Fock space.

The degeneracies of the BPS states corresponding to this gas of spinning M2 branes in M-theory on $X\times R^{4,1}$ are computed by the topological string on $X$ \cite{GV1,GV2}.
The M2-brane particles on $X$ are specified by the class $\beta\in H_2(X)$ which M2-brane wraps, and the spins under the $SO(4)=SU(2)_L\times SU(2)_R$ little group of a massive particle in 5d.
We denote the number of M2-brane particles with $intrinsic$ spin $(m_L, m_R)=(2j_L,2j_R)$ and wrapping $\beta$ by $N_{\beta}^{(m_L,m_R)}$, and define the index
$$
N_{\beta}^{m}=\sum_{m_R} (-1)^{m_R} N_{\beta}^{(m_L,m_R)}.
$$
For each such particle, we get a field $\Phi$ in 5D, which can carry additional excitations on $R^4$. If we denote by $z_1,z_2$ the coordinates on $R^4$, $\Phi$ has a mode expansion
$$
\Phi(z_1,z_2) = \sum_{i, j} \alpha_{i,j} z_1^i z_2^j.
$$
As noted in \cite{DVVafa}, $z_1$ and $z_2$ transform under the $U(1) \subset SU(2)_L$ with charge one each --- so the excitations on $R^4$ also carry D0 brane charge.

Each 5D particle corresponding to M2 brane in class $\beta$ and intrinsic spin $m$ thus contributes a factor
$$
\prod_{i,j =1}(1-q^{i+j+m-1} Q^{\beta})
=\prod_{n=1}(1-q^{n+m} Q^{\beta})^{n},
$$
to the generating function,where $q$ keeps track of the spin and $Q$ of the class in $H_2(X)$.
The contribution from all particles summed up is therefore given by
\beq
\scZ^c_{\rm Fock}=\prod_{\beta,m} \prod_{n=1}^{\infty}
(1-q^{l+m} Q^{\beta})^{lN_{\beta}^m}.
\label{closedZFock}
\eeq
Note that the product \eqref{closedZFock} involves both $\beta>0$ and $\beta<0$.
For particle corresponding to class $\beta$, we also get its anti-particle, corresponding to the M2 brane wrapping $\beta$ with opposite orientation.

This amplitude is computed by the topological string partition function on $X$ \cite{GV1,GV2}
\beq
\scZ^c_{\rm top}=\prod_{\beta>0,m} \prod_{n=1}^{\infty}
(1-q^{n+m} Q^{\beta})^{nN_{\beta}^m},
\eeq
where $Q$ counts the degree of the worldsheet instanton, and $q=e^{-g_s}$. Note that this only receives contribution from M2 branes on holomorphic curves, and not anti- M2 branes. Putting this together, the Fock space partition function is
\beq
\scZ^c_{\rm Fock}=\scZ^c_{\rm top}(q,Q) \scZ^c_{\rm top}(q,Q^{-1}).
\eeq

We still have to restrict the Fock space to be generated by mutually BPS particles. This is determined by the central charges in four dimensions, and depends on the chamber. By assumption 2, we are choosing all central charges to be aligned or anti-aligned, corresponding to turning off the K\"ahler class, and allowing for only the B-fields in IIA.  The central charge of D2 brane in class $\beta$ with $n$ D0 branes is
$$
Z(D2)=(\beta B+n)/R.
$$
The BPS particles are those which satisfies
$$
Z(D2)>0.
$$
This means that the BPS partition function is given by
$$
\scZ^c_{\rm BPS}=\scZ^c_{\rm top}(q,Q) \scZ^c_{\rm top}(q,Q^{-1}) \Big|_{\rm chamber}.
$$
The fact that the particles span a free Fock space, and the partition function takes an infinite product form, is precisely what is observed in the literature \cite{Szendroi,Young1,Nagao1}.

\section{Open BPS States and M-theory}\label{sec.opentop}

In this section, we consider the generalization of the above to open BPS invariants\footnote{See \cite{Nagao2,NY,Nagao3} for recent discussions.}.
Consider adding $M$ D4 branes wrapping a special Lagrangian submanifold
$L$ of $X$. The D4 branes fill $R^{1,1}$ subspace of the flat space, and break half the supersymmetry of the Calabi-Yau. The theory on the branes thus has ${\cal N}=(2,2)$ supersymmetry in
two dimensions. We now get new kinds of BPS particles, corresponding to D2 branes wrapping disks and ending on the D4 brane \cite{OVknot}. These particles can form BPS bound states with the D6 brane and the closed D2 branes and the D0 branes we had before, and pin them to the D4 branes. These are the open BPS invariants that we would like to count\footnote{The counting of open BPS states was
also studied in \cite{ANV}. In that context, the instead of the D6 branes wrapping $X$ the authors had $N$ D4 branes on a divisor in $X$.}.

A D2 brane ending on a D4 brane is magnetically charged under 5d gauge field $A$ on the D4 brane, and electrically charged under the corresponding dual two-form $B$, $dA =*dB$. Suppose $b_1(L) = r$. We get $r$ $1-$cycles on $L$ that are contractible in $X$ and fill in to holomorphic disks in $X$. We will assume, for simplicity that $r=1$, tough the generalization to arbitrary $r$ is manifest. Integrating the B-field on the $S^1$ we get a  $U(1)^M$ magnetic gauge group generated by $\int_{S^1} B$. The D2 branes wrapping the holomorphic disks and ending on the D4 brane
are BPS particles charges under the magnetic $U(1)^M$. The $U(1)^M$ gauge fields in two dimensions sit in the twisted chiral multiplets $\Sigma_i$, $i=1, \ldots M$, whose lowest components $u_i$ enter the BPS masses of the particles charged under them\footnote{In fact, as we will need later, $u = \int_{\rm disk} k + i \int_{S^1} A$, where $A$ is the electric $U(1)$ on the D4 brane. We can write this, equivalently as $u = \int_{\rm disk} (k + iB_{NS}),$ since on the D4 brane $B_{NS} - dA$ is the gauge invariant combination.}. Since we are not interested in the gauge dynamics on the D4 branes, we will view the gauge symmetry as a global symmetry, and $\Sigma_i$ as background multiplets. The charges of a D2 brane under $U(1)^M$ keep track of which D4 branes the D2 brane has boundaries on, and how many times it wraps the corresponding $S^1$.
More precisely, since the $M$ D4 branes are identical, we have $S_M$ permutation invariance so the particles are representations $R$ of $U(1)^M/S_M$ (which can also be viewed as representations of $U(M)$).

When we lift this to M-theory on $X \times TN\times R$, we now get an M5 brane wrapping the Lagrangian $L$
and filling $R^{2,1}$, where open D2 branes become M2 branes ending on the M5 brane. On the M5 branes, there is a $U(1)^M$ gauge theory, with ${\cal N}=2$ supersymmetry, where the open M2 branes are BPS particles.
Under the same assumptions as in the previous section (we will explain below why these are justified), the computation of BPS bound states in IIA reduces to computation of BPS degeneracies of a gas of free particles in M-theory, now living on the flat $R^{2,1}$ world volume of the M5 brane.
Below, we will show,  following  \cite{OVknot}, that the degeneracies of the spinning open M2 branes on $R^{2,1}$ are computed by the open topological
string, corresponding to A-model on $X$ with a Lagrangian $L$. To extract the degeneracies of the BPS bound state of one D6 brane with the open and closed D2 branes and D0 branes, we have to restrict the Fock space to those states preserving the same supersymmetry.

\subsection{Open M2 Branes and the Topological String}

The BPS particles are now labeled by their $U(M)$ representation $R$, bulk class $\beta$, spin $s$ and $R$-charge $r$.
The presence of the M5 branes breaks the symmetries: $SO(4) = SU(2)_L\times SU(2)_R$ is now broken to $SO(2)_L\times SO(2)_R$: these now correspond to the little group of a particle in three dimensions and the R-symmetry of the ${\cal N}=2$ theory. More precisely $s_L = s+r, s_R = s-r$ where $s$ is the spin and $r$ is the R-charge. $s_R$ permutes the fields within the same multiplet, and $s_L$ annihilates the state.

We denote the number of M2-brane particles with $intrinsic$ spin $(s_L, s_R)$ and wrapping $\beta$ and representation $R$ by $N_{\beta, R}^{(s_L,s_R)}$, and define the index
$$
N_{\beta, R}^{s_L}=\sum_{s_R} (-1)^{s_R} N_{\beta, R}^{(s_L,s_R)}.
$$
Now, again, each such 3d particle gives rise to a field $\Phi$, and excitations of this field on $R^2$ are the particles we want to count
$$\Phi(z) = \sum_{n} \alpha_n z^n.
$$
Here we have set $z=z_1$, since M5 brane wraps the and $z_2=0$ subspace of $R^{4,1}$.

The partition function of these particles can be written as follows.  Introduce chemical potentials $q$, corresponding to the spin $s$, $Q$ corresponding to the charge $\beta$ and $v_i$
corresponding to the charges of the M2 branes under the $U(1)^M$.  For each field in representation $R, \beta, s$, we get a
contribution
$$
\prod_{{\vec k}_R}\prod_{n=1} (1- q^{s+n} Q^\beta  v^{\vec k_R})^{m_{{\vec k}_R}},
$$
where the product is over the weight vectors  ${\vec k}_R$ are of the representation $R$, ${m_{{\vec k}_R}}$ are the corresponding multiplicities and
$$v^{\vec k_R}=\prod_{i=1}^M v_i^{k_{R_i}}.$$
The full Fock-space partition function is
\beq
\scZ^c_{\rm Fock}=\prod_{\beta, s, R} \prod_{k_R} \prod_{n=1}^{\infty}
(1-q^{s+n}v^{\vec k_R} Q^{\beta})^{{m_{{\vec k}_R}} N_{\beta,R}^s}.
\eeq
Note that, as in the closed string case, this should include M2 branes with both orientations. Flipping the orientation corresponds to sending $\beta$ to $-\beta$, $s$ to $-s$ and $R$ to ${\bar R}$, simultaneously, where CPT ensures
\beq
N_{\beta,R}^s = N_{-\beta,{\bar R}}^{-s}\, .\label{anti}
\eeq

As explained in \cite{OVknot}, the degeneracies of open M2 branes ending on an M5 brane wrapping $L$ are computed by the open topological string partition function in the presence of a D4 brane wrapping a Lagrangian $L$.
It was shown in \cite{OVknot} that the open topological string partition function has the following simple expansion, similar to the Gopakumar-Vafa expansion \cite{GV1,GV2} in the closed case:
\beq
\scZ^o_{\rm top}=\exp\left(
\sum_{d=1}^{\infty} \sum_R f_R(q^d,Q^d) \textrm{Tr}_R \frac{V^d}{d}
\right),
\label{OVexpansion}
\eeq
where
$$
f_R(q,Q)=\sum_{s,\beta>0}\sum_R \frac{N_{R,\beta,s}}{q^{1/2}-q^{-1/2}} Q^{\beta} q^{s-{1\over 2}},
$$
and where  $N_{R,\beta,s}$ is an integer counting M2-brane particles with $R,\beta$ and $s$\footnote{We have shifted the definition of spin by $1/2$ relative to \cite{LMV}. }.
The symbol $\textrm{Tr}_R V$ denotes the holonomy of the gauge field on the D4-brane. Here $V$ captures the BPS masses of the open M2 branes, namely
$$
V=\textrm{diag}(v_1,v_2,\ldots,v_M).
$$
We can write
\beq
\textrm{Tr}_R V^d=\sum_{k_R} {m_{{\vec k}_R}} \prod_i v_i^{d k_i^R},
\label{TrRV}
\eeq
where $k_i^R$ are the weights of the representation $R$ and $m_{{\vec k}_R}$ their multiplicities.
Then expression \eqref{OVexpansion} can be rewritten as the Fock space trace:
\beq
\scZ^o_{\rm top}=\prod_{l=0}^{\infty}\prod_{R,s,\beta>0} \prod_i \left(1-q^{n+s} Q^{\beta} v^{k^R} \right)^{{m_{{\vec k}_R}}N_{R,\beta,s}}.
\eeq
This naturally corresponds to the half of the Fock space corresponding to M2 branes wrapping holomorphic curves only. In other words, we have
\beq
\scZ^o_{\rm Fock}=
\scZ^o_{\rm top}(q,Q,v) \scZ^{o,*}_{\rm top}(q,Q^{-1},v^{-1}),
\label{openZsquare}
\eeq
where $\scZ^{o,*}_{\rm top}$ contains the contributions of the anti-M2 branes,
\beq
\scZ^{o,*}_{\rm top}(q,Q^{-1},v^{-1})=\prod_{l=0}^{\infty}\prod_{R,s,\beta>0} \prod_i \left(1-q^{n-s} Q^{-\beta} v^{-k^R} \right)^{{m_{{\vec k}_R}}N_{R,\beta,s}},
\eeq
where we used the \eqref{anti} and the fact that $\textrm{Tr}_{{\bar R}}V = \textrm{Tr}_{R} V^{-1}$. Note that
$\scZ^{o,*}_{\rm top}(q,Q,v)$ is almost the same as $\scZ^{o}_{\rm top}(q,Q,v)$, the only difference being that the spin $s$ in the power of $q$ is replaced by $-s$.

\subsection{Open BPS Wall Crossing and M-theory}\label{sec.openwc}

Up to this point we discuss all M2-brane particles. Consider the central charges of the particles, which determine which ones are mutually BPS.
The central charge of the open D2 brane wrapping a holomorphic disk in class $k_R$, bulk class $\beta$ and D0 brane charge $n$
is given by\footnote{As in \cite{AOVY}, there is a complex proportionality constant between $R$ in the formulas here and the radius of the Taub-NUT; this is irrelevant for the discussion of wall crossing.}
\beq\label{cc}
Z(D2)=\left(u(k_R)+t(\beta) +n\right)/R,
\eeq
where
$$
t = \int_{S^2} i k + B,
$$
and
$$
u= \int_{\rm disk} ik + B.
$$
Here, $B$ is the NSNS B-field, and $k$ is the K\"ahler class and $R$ is the radius of the Taub-NUT defined above. The central charge receives contributions from the background twisted chiral multiplets where gauge fields and their superpartners reside (see, for example \cite{HoriHanany}). The central charge is of the form $\sigma(q) = q \cdot \sigma$ where $q$ is the charge of the state under the gauge symmetry, and $\sigma$ the lowest component of the twisted chiral multiplet. The first term in \eqref{cc} comes from the $U(1)^M$ gauge symmetry on the D4 branes. The second two come from the bulk gauge fields, reduced to two dimensions\footnote{The 4d ${\cal N}=2$ bulk vector multiplet splits into two twisted chiral multiplets. These determines the 2d ${\cal N}=(2,2)$ central charges of the 4d electrically and magnetically charged states.  Namely, the twisted chiral multiplet coupling to electrons contains the the $d=4$ complex scalar, and the longitudinal components $A_{0,1}$ of the $d=4$ vector. The twisted chiral multiplet coupling to monopoles is slightly more complicated. Start with the chiral multiplet containing two transverse components of the gauge fields $A_{2,3}$. The dual twisted chiral multiplet determines the $d=2$ central charge of the magnetically charged particles, in particular of the D6 brane. We will need this in a later section.}.
To satisfy our assumption 2, that all the central charges of D2 and D0 branes align, we had set the area of all the disks to zero and 2-cycles to be zero. This is possible, since $t$ and $u$ are  the moduli of our our theory in two dimensions, which we can dial at will.
The central charge is then given by
$$
Z(D2)=\left(B(k_R)+ B(\beta)+{n}\right)/R,
$$
in IIA, or equivalently
$$
Z(M2)=C(k_R)+C(\beta)+{n/R},
$$
in M-theory.
The BPS particles are particles whose central charges align, with
$$
Z(D2)>0.
$$
This means that the complete open+closed BPS partition function is
$$
\scZ^{o+c}_{\rm BPS}(q,Q,v)=\scZ^{o+c}_{\rm Fock}(q,Q,v) \Big|_{Z(D2)>0},
$$
where the Fock-space partition function is computed by the open+closed topological string partition function,
$$
\scZ^{o+c}_{\rm BPS}(q,Q,v)=\scZ^{o+c}_{\rm top}(q,Q,v) \scZ^{*,o+c}_{\rm top}(q,Q^{-1},v^{-1})  \Big|_{Z(D2)>0}.
$$


\subsection{Examples}\label{sec.examples}

In \cite{NY}, a crystal melting model for the open BPS invariants at hand was proposed by one of the authors, generalizing the previous results for the closed invariants \cite{OY1,OY2}. There the jump of the open BPS invariants under the change of closed string moduli was studied. However,in that work it was not clear which value of the open string moduli the crystal computation corresponded to, or what happens as open string moduli are varied.
The partition function that the authors computed had two properties: a) in the limit large
values of closed K\"ahler moduli it reduced to a single copy of open topological string, and b)
as the closed string moduli are varied, only the degeneracies of closed BPS states jumped.

From our results, property a) implies that BPS invariants studied in \cite{NY}
are in the special chambers where
$$R>0, \qquad u^i \to\infty.$$
Then, fixing $R$, and $u^i$ and taking $\textrm{Re}(t)$ to infinity one indeed recovers the topological string amplitude.
For example, when $0<\textrm{Re}(t)<1$, we have
$$
\scZ_{\rm BPS}=\scZ^{o+c}_{\rm top}(q,Q,v) \scZ^c(q,Q^{-1}).
$$
and when $\textrm{Re}(t)\to\infty$, we have
$$
\scZ_{\rm BPS}=\scZ^{o+c}_{\rm top}(q,Q,v).
$$
In this case, taking $\textrm{Re}(t)$ to infinity in addition, only M2 branes ever contribute, and no anti-M2 branes. This means that $\scZ_{\rm BPS}$ has the same open part as the topological string theory.

We can consider another extreme $0<\textrm{Re}(u^i)<1$. In this case, open part has contributions both from M2-branes and anti M2-branes. For example if $0<\textrm{Re}(t)< 1$
$$
\scZ_{\rm BPS}(q,Q,v)=\scZ^{o+c}_{\rm top}(q,Q,v) \scZ^{o+c,*}_{\rm top}(q,Q^{-1},v^{-1} ).
$$
We can also consider $R<0$. For example, there is a chamber $R<0, \textrm{Re}(u^i)\to\infty, \textrm{Re}(t)\to \infty$, where the partition function simply becomes one.


\section{Relation to the Work of Cecotti and Vafa}

The wall crossing of the closed BPS invariants discussed in section 2 is a special case of a more general problem, the wall crossing of BPS bound states in four dimensional theories with ${\cal N}=2$ supersymmetry.
In this context, Kontsevich and Soibelman \cite{KontsevichS} recently conjectured that the degeneracies on two sides of the wall are related by commuting certain symplectomorphisms of complex tori. In the special case of ${\cal N}=2$ gauge theories in four dimensions, this remarkable structure was explained from several different perspectives \cite{GMN1,GMN2, CecottiV}. In particular, in \cite{CecottiV} both the appearance, and the particular choice of symplectomorphism, was beautifully illuminated. In some cases, the statements that follow from \cite{KontsevichS} are particularly simple, and were predicted in physics literature \cite{DenefM} a while back. When the state of charge $\Gamma$
decays into two primitive states $\Gamma_1$ and $\Gamma_2 = \Gamma-\Gamma_1$,
$$
\Gamma\; \rightarrow \; \Gamma_1 + \; \Gamma_2,
$$
the degeneracies of single particle states $\Omega(\Gamma)$ jump\footnote{In what follows, we will be cavalier about whether we gain or loose the states in the jump, i.e. about the sign of $\Delta\Omega$, since that depends on the direction of crossing.} as \cite{KontsevichS, DenefM}
$$
\Omega(\Gamma) \rightarrow \Omega(\Gamma) + \langle\Gamma_1, \Gamma_2\rangle \;\Omega(\Gamma_1)\;\Omega(\Gamma_2).
$$
%
More generally, when the central charges of $\Gamma_1$ and $\Gamma_2$ align, they do so for any multiple of $\Gamma_1$ and $\Gamma_2$ as well, so one can a-priori loose an infinite number of states. While there are no known such simple formulas for the general case, \cite{KontsevichS,DenefM} predicted that in the semi-primitive case, where only one of the charges is primitive, the degeneracies $\Omega(\Gamma_1+n \Gamma_2)$
jump as
%
%
$$
Z(q) \rightarrow  Z(q)       \;\prod_{n=1}^{\infty}(1-q^n)^{-n \langle\Gamma_1, \Gamma_2\rangle\Omega(n\Gamma_2)},
$$
where we defined
\beq\label{zq}
Z(q)=\sum_{n=0}^{\infty} \Omega(\Gamma_1+n \Gamma_2) q^n.
\eeq
As noted in \cite{AOVY}, M-theory provides a derivation of this formula when $\Gamma_2$ carries
D0 and D2 brane charges, and $\Gamma_1$ in addition carries one unit of D6 brane charge. Note that the jumps are governed by a Mac-Mahon function type formula, $\prod_{n=1}^{\infty}(1-q^n)^{-n \omega(n)}$.

In the present context, we count open BPS states. We have seen that
M-theory predicts that the BPS states jump as
\beq\label{jump}
\Omega(\Gamma) \rightarrow \Omega(\Gamma) + \Omega(\Gamma_1)\;\Omega(\Gamma_2).
\eeq
in the primitive case, and more generally as,
\beq\label{jumptwo}
Z(q) \;\;\rightarrow \;\;Z(q) \; \prod_{n=1}^{\infty}(1-q^n)^{-\Omega(n\Gamma_2)},
\eeq
in the semi-primitive case, where $Z(q)$ is as in \eqref{zq}. In other words, M-theory predicts that the open BPS states jumps are governed by eta-function $\prod_{n=1}^{\infty}(1-q^n)^{-1}$ type formulas.

In fact, this could have been anticipated. The open BPS states are BPS states in the ${\cal N}=(2,2)$ abelian gauge theory in two dimensions on the world-volume of the D4 branes. The open D2 branes (bound to the closed D0, D2 branes and the D6 brane), as discussed above, are charged particles in this two-dimensional theory.
The particles in two dimensions are solitons. This is so even for fundamental matter, simply because a particle is a codimension one object, so the vacuum of the theory can change from one side to the other. Soliton spectra of the massive ${\cal N}=(2,2)$ theories were studied in \cite{CFIV, CecottiVold}. The fact that the spectrum of BPS states can jump was in fact discovered in this context. We will now first review the essential results from \cite{CecottiVold}, and then explain how to apply them in the present context.

\subsection{Review of \cite{CecottiVold}}
Consider a massive ${\cal N}=(2,2)$ theory.
BPS particles are solitons $\Delta_{ik}$ interpolating between the vacua $i$ and $k$ at spatial infinities. Let
$$
\mu_{ik}={\rm Tr}_{ik}  (-1)^F {\rm F}
$$
be the ``number" of such solitons, or more precisely the index  that weights the solitons with their fermion number charge $F$. Only the BPS solitons contribute to the index. These live in short, two dimensional representations of the ${\cal N}=(2,2)$ supersymmetry, and the index is the same as counting the multiplets weighted by $(-1)^F$, where $F$ is the fermion number of the lowest component in the multiplet \cite{CecottiVold}.

The central charge, $Z(\Delta_{ik})$ of the soliton depends only on the vacua $i$ and $k$,
so we can define $W_{i}$, so that
$$Z(\Delta_{ik}) = W_i - W_k.
$$
If the theory at hand is a Landau-Ginsburg theory, then $W_i$ is the value of the superpotential $W$ in the $i$-th vacuum.
If, as we cross the line of
    marginal stability, $W_j$ crosses the line in the $W$-plane interpolating between $W_i$ and $W_k$, the soliton $\Delta_{ik}$ decays as
$$\Delta_{ik}\rightarrow \Delta_{ij}+\Delta_{jk},$$
and moreover the number of solitons jumps as
%
\beq\label{cvone}
\mu_{ik} \rightarrow \mu_{ik} + \mu_{ij}\mu_{jk}.
\eeq
More generally, vacua $j_1, j_2, \ldots j_N$ can cross the straight line connecting $W_i$ and $W_j$, where the order is set by the order of interception points. The state $\Delta_{ik}$ can have different decay channels, corresponding to all the different ways of getting from $i$ to $k$ while passing through the intermediate vacua.
The index in the $ik$ sector jumps as
\beq\label{cvtwo}
\mu_{ik} \rightarrow \mu_{ik} + \sum_{\stackrel{1\leq n \leq N}{1\:\leq s_1 <s_2 <\ldots< s_n \leq N}}
\mu_{ij_{s_1}}\mu_{j_{s_1},j_{s_2}}\ldots \mu_{j_{s_n}, k}.
\eeq
This has a simple interpretation. The jump corresponds to counting all the chains of BPS solitons interpolating from vacuum $i$ to vacuum $k$ via the intermediate vacua.
More precisely, we are counting the lowest components of these multiplets, weighted by $(-1)^F$. Since the net fermion number is simply the sum of the fermion numbers, the result is simply the product of the BPS degeneracies in the individual sectors of the chain.

\subsection{Counting of Open BPS States}

In the case at hand, we are counting massive BPS particles, which carry charges under the $U(1)$ gauge fields from the D4 brane and the bulk.
%
%
The corresponding central charge that enters the ${\cal N}=(2,2)$ supersymmetry algebra is
determined by the lowest components $\sigma_{\alpha}$ of the
twisted chiral multiplets $\Sigma_{\alpha}$\footnote{In this section, for simplicity of the notation, we use $\sigma$ to denote both open and closed twisted masses of the particles on the D4 branes. These include the central charges of D0, D2 and D6 branes, bound to open D2 branes. In addition, since we are not interested in the dynamics of the gauge theory on the D4 branes, but only in the BPS particle content, we have been viewing $\Sigma_{\alpha}$ as non-dynamical, so that the charges $q^{\alpha}$ are simply the global symmetry charges.}. 
For a particle $\Gamma$
of charge $q^{\alpha}$, the central charge is
\beq\label{cc2}
Z(\Gamma) = \sum_{\alpha} q^{\alpha}  \sigma_{\alpha}.
\eeq
Per definition then, the solitons of the theory are the fundamental particles. Consider a vacuum of the theory $|i\rangle$. Adding a BPS particle $\Gamma$ to it changes the vacuum of the theory from $|i\rangle$ on side to $|j\rangle$ on the other where, again per definition, 
$$
W_j - W_i = Z(\Gamma).
$$
We can think of the vacua as {\it labeled by the charges of the particles needed to create it}, from an arbitrary but fixed vacuum. Note that we have extra structure here, that is not present in an arbitrary massive ${\cal N}=(2,2)$ theory.  Namely, given any vacuum $|i'\rangle$, the BPS particle $\Gamma$ takes it to a vacuum $|j'\rangle$ where $W_{j'}$ and $W_{i'}$ differ by the BPS mass of $\Gamma$,  $W_{j'} - W_{i'} = Z(\Gamma)$.

\begin{figure}[htbp]
\centering{\includegraphics[scale=0.6,trim=0 360 0 80]{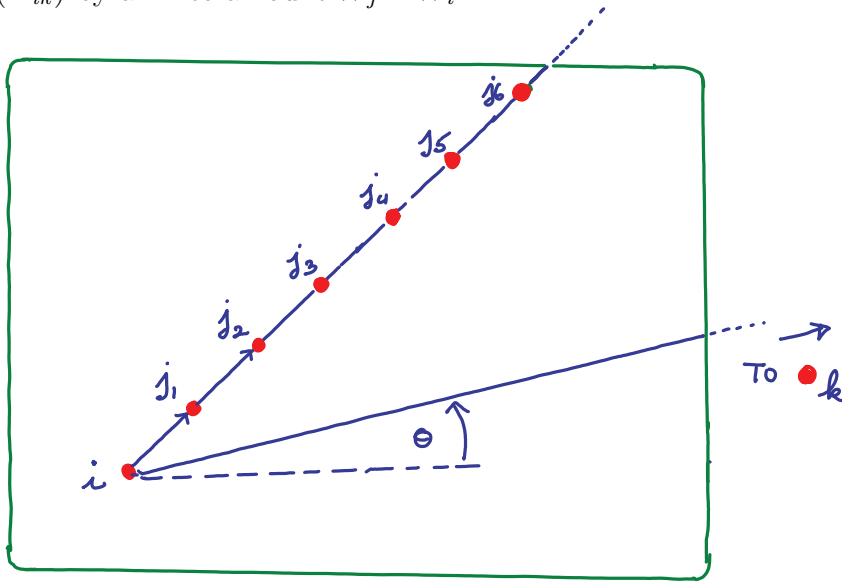}}
\caption{In the $W$-plane, all the $W_{j_i}$'s are on a single line emanating from $W_i$. $W_k$ is infinitely large, and we cut off the $W$-plane at large radius $\Lambda$, shown here in green.}
\label{W-plane}
\end{figure}

Among the particles in the theory is a D6 brane wrapping the whole non-compact Calabi-Yau. The focus of our paper has been understanding the wall crossing of particles carrying one unit of D6 brane charge. However, the corresponding central charge is strictly speaking infinite. If $|k\rangle$ is a vacuum with one unit of D6 brane charge, then $W_k$ is strictly at infinity in the $W$ plane.
To deal with this infinity we need a regulator. The correct way to do this, it turns
out, is to cut off the W-plane at some large but finite radius $\Lambda$. $\Delta_{ik}$ is then a non-compact cycle with a boundary. The relevant data that affects the problem is only the angle $\theta$ that $W_k-W_i$ makes with the real axis in the $W$-plane\footnote{This parallels the treatment in \cite{JM} where the wall crossing of closed BPS bound states corresponding to D6 branes on non-compact Calabi-Yau was studied.} (see Fig. \ref{W-plane}). Moreover, it is easy to see that $\theta$ is independent of any additional D2 or D0 brane charges that the D6 brane may carry, since the central charges of these are finite. So, any soliton $\Delta_{jk}$
where $W_j$ is in $|W| \leq \Lambda$, corresponds to a state with the same value of $\theta$, but whose central charge $Z(\Delta_{jk})$ differs from $Z(\Delta_{ik})$ by a finite amount $W_j-W_i$.

Fix now $\Gamma = \Delta_{ik}$, corresponding to one unit of D6 brane charge, and some D2 brane and D0 brane charges. The degeneracy $\mu_{ik}$ of domain walls interpolating between the vacua $i$ and $k$ is simply the degeneracy of the 1-particle states with charge $\Gamma$,
$$\mu_{ik} = \Omega(\Gamma).$$
As we vary the moduli, suppose vacuum $j$ inside the cutoff $W$-plane crosses the straight line between $i$ and $k$. This then implies that corresponding state $\Gamma_2$ is a single particle state, with some D0 and D2 brane charges, and no D6 brane charge. So, $\Delta_{ij} = \Gamma_2$, and $\Delta_{jk} = \Gamma-\Gamma_2=\Gamma_1$ carries one unit of D6 brane charge. The soliton degeneracies are simply the degeneracies of the corresponding one-particle states, so
$$\mu_{ij} = \Omega(\Gamma_2),\qquad\mu_{jk} = \Omega(\Gamma_1),$$
and the jump \eqref{cvtwo} in the degeneracies of $\Gamma$ predicted from \cite{CecottiVold} agrees with \eqref{jump}.

Moreover, if the vacuum $|j\rangle$ corresponding to $\Gamma_2$ crosses the $ik$ line in the $W$ plane, a large number of other vacua will cross as well.
These are vacua which can be obtained from $|i\rangle$ by adding $n$ $\Gamma_2$ particles. They are all collinear, located at points $W_{j_n} = W_i + n Z(\Gamma_2)$. This holds for any $n$, as long as $W_{j_n}$ is inside the cut-off W-plane. These additional critical points lead to more complicated decays.
Let's fix $\Gamma_1=\Delta_{j_n k}$, the decay product with one unit of D6 brane charge, and consider all the different ways that we can obtain it from $\Gamma = \Delta_{ik}$. There is is a two-soliton decay,
$$
\Delta_{ik} \rightarrow \Delta_{i j_n}+\Delta_{j_n k}.
$$
where $\Gamma=\Gamma_1+n \Gamma_2$
decays into $\Gamma_1$ and $n \Gamma_2 =\Delta_{i j_n}$
This, according to \cite{CecottiVold} corresponds to a change in the degeneracies
$$
\mu_{ik} \rightarrow \mu_{ik} + \mu_{ij_n}\mu_{j_nk},
$$
or
$$
\Omega(\Gamma)  \rightarrow \Omega(\Gamma)+ \Omega(\Gamma_1) \Omega(n\Gamma_2).
$$
This is exactly as predicted from M-theory.

There are also channels where $n\Gamma_2$ is split into more particles, each carrying some multiple of $\Gamma_2$ charge. Here we need to be careful. Much of the discussion of \cite{CecottiVold} assumes that no three vacua are collinear in the $W$-plane\footnote{If collinearity is coincidental, one can remove it without loss of generalities, simply by displacing the vacua slightly. However in the present case it is physical, and there is no small deformation of the theory that can remove it.}. Here, we are in the opposite regime, of a large number of collinear vacua. In this regime, a very naive application of \eqref{cvtwo} overcounts the change in the soliton number.

Recall that, from \eqref{cvtwo}, the number of solitons gained or lost, is $\Omega(\Gamma_1)$ times the number of ways of assembling a chain of solitons of total charge $n\Gamma_2$ out of the available $\Omega(k \Gamma_2)$ one particle BPS states particles of charge $k\Gamma_2$, where $k=1,2 \ldots $. In counting the number of possibilities, the order of in which we string the $\Gamma_2$-solitons in a chain would matter. When the vacua are not collinear, the order indeed matters. However here this manner of counting would lead to a contradiction.
Consider example, in a decay channel where $\Gamma$ splits into $\Gamma_1$ and $n$ copies of $\Gamma_2$,
$$
\Delta_{ik} \rightarrow  \Delta_{ij_1}+\Delta_{j_1 j_2}+\ldots +\Delta_{j_nk},
$$
The $n$ $\Gamma_2$ particles are all mutually BPS, so in counting the number of such $n$ particle states, we can simply use the free Fock-space, generated by $1$ particle states. The dimension of the corresponding $n$-particle Hilbert space is not $\Omega(\Gamma_2)^n$, as the naive application of \cite{CecottiVold} would suggest, but $\Omega(\Gamma_2) (\Omega(\Gamma_2)+1)\ldots
(\Omega(\Gamma_2)+n-1)/n!.$ The number of states we gain or loose is just this times $\Omega(\Gamma_1)$. One should in principle be able to verify this by a careful computation of $n$-soliton contributions to the index in \cite{CFIV, CecottiVold}. More generally,
we can consider more complicated splits, where $n$ units of $\Gamma_2$ charge are split among $m_{\ell}$ particles of charge $\ell \Gamma_2$, with $\sum_{\ell} m_{\ell} \ell = n$. The number of states is just the product of the dimensions of $m_{\ell}$ particle Hilbert spaces. 
Of course, this is precisely the counting predicted from M-theory.

\section*{Acknowledgments}
M.~Y. would like to thank D.~Krefl and K.~Nagao for collaborations on related projects.

M.~Y. is supported by JSPS Fellowships for Young Scientists.
M.~Y. is also supported by DOE grant DE-FG03-92-ER40701, by the JSPS fellowships for Young Scientists, by the
World Premier International Research Center Initiative, 
and by the
Global COE Program for Physical Sciences Frontier at
the University of Tokyo, both by MEXT of Japan.
M.Y. would also like to thank Berkeley Center for Theoretical Physics for hospitality, where part of this work has been performed.

\bibliographystyle{JHEP}
\bibliography{dthesis}

\end{document}